\documentclass[a4paper,12pt]{article}
\usepackage{latexsym}
\usepackage{amsmath}
\usepackage{amssymb}
\usepackage{graphicx}
\usepackage{wrapfig}
\usepackage{fancybox}
\usepackage{bm}
\def\id{{\rm d}}

\def\be{\begin{equation}}
\def\ee{\end{equation}}
\def\bea{\begin{eqnarray}}
\def\eea{\end{eqnarray}}
\def\nn{\nonumber\\}
\begin{document}
\setlength{\baselineskip}{8mm}

%
\begin{center}
\setlength{\baselineskip}{8mm}
{\bf\Large\sf
Long-range and selective coupler for superconducting flux qubits
}

\vspace{2em}

{\large\bf\sf  Hayato Nakano,  Kosuke Kakuyanagi, }

\vspace{1em}
{\small\it NTT Basic Research Laboratories, NTT Corporation},  {\it Atsugi-shi, Kanagawa 243-0198, Japan}
\vspace{1em}

{\large\bf\sf Masahito Ueda}
\vspace{1em}

{\small\it Department of Physics, Tokyo  Institute of Technology}, {\it Meguro-ku, Tokyo 152-8551, Japan}

{\it NTT Basic Research Laboratories, NTT Corporation},  {\it Atsugi-shi, Kanagawa 243-0198, Japan}
\vspace{1em}

and {\large\bf\sf  Kouichi Semba}
\vspace{1em}

{\small\it NTT Basic Research Laboratories, NTT Corporation},  {\it Atsugi-shi, Kanagawa 243-0198, Japan}

\end{center}

\vspace{2mm}

\setlength{\baselineskip}{8mm}

\setlength{\baselineskip}{8mm}

{\Large\bf Abstract}

We propose  a qubit-qubit coupling scheme for superconducting flux quantum bits (qubits), where 
a quantized Josephson junction resonator and microwave
irradiation are utilized.  The junction is used as a tunable inductance  controlled by changing the bias current flowing through the
junction, and thus the circuit works as a tunable resonator.
This enables us to make any qubits interact with the resonator. 
Entanglement between two of many qubits
whose level splittings satisfy some conditions,
is formed by microwave irradiation causing a two-photon Rabi oscillation.
Since the size of the resonator can be as large as sub-millimeters and qubits interact with it via mutual inductance, our scheme 
makes it possible to construct  a quantum gate involving remote qubits. 


\newpage

One of the major challenges in solid-state quantum computation
is to selectively couple two among many qubits that are located at different places in the system.
In this letter, we present a scheme to overcome this difficulty.


A practical qubit-qubit coupler should allow, 
(i) coupling between qubits can be switched on/off
at will; (ii) coupling that is strong enough so that the desired entanglement can be
formed within the coherence time of the qubit system,
(iii) selective coupling between two arbitrary qubits
among many qubits, (iv) coupling between two remote qubits.

For superconducting qubits, various coupling schemes have been proposed.
Some of them add the third element $^{1-6}$  
to achieve controllability in  two-qubit coupling.

Schematical original idea has appeard in Ref. 1. 
Bias current control of the coupling
was shown in Ref. 2. 
You {\it et al.} $^3$ 
proposed a two-qubit gate with a superconducting LC circuit.
Its mechanism is an analogue of 
the Cirac-Zoller scheme originally proposed for atomic or
ionic qubit systems $^7$. 
A DC-SQUID is used as the medium for the interaction in the proposal by 
Plourde $^4$ 
where the SQUID is used as a  flux-transformer controlled by changing an external magnetic field and applied DC bias current. 
Bertet {\it et al.}  $^5$ improved the controllability of the SQUID flux-transformer by introducing AC bias current into
the SQUID.
A third qubit is introduced as the medium of two qubits
in the proposal by Niskanen et al. $^6$. 

Above schemes $^{3-6}$  
satisfy the requirments (i) and (ii).
One $^3$ 
can be used a qubit-qubit coupler allowing us to couple 
spacially separated qubits, that is, to satisfy requirement (iv).
In particular, requirement (iv) is important when we integrate many flux-qubits on the
same chip  using microfabrication techniques,
in order to directly couple two target qubits that do  not neighbor each other.


The qubit-qubit coupling scheme we proposed here for superconducting flux qubits satisfies
all of the above requirements. We use a tunable resonator and two-photon Rabi oscillation. 
The resonant frequency of a circuit consisting of  capacitance $C$ and  inductance $L$ is given by
\be
\omega_{\rm r}=\frac1{\sqrt{L C}}=\frac1{\sqrt{(L_0+L_{\rm J}) C}},
\label{wa}
\ee
where we put $\hbar=1$.
$L_0$ is the geometrical and kinetic inductance of the circuit, and $L_{\rm J}$ stands for 
the Josephson inductance, $L_{\rm J}=(\frac1{2e})^2/E_{\rm J}$ where $E_{\rm J}$ is the Josephson energy of the junction.
The effective Josephson energy, i.e., the curvature at the bottom of the Josephson potential 
varies with the DC bias  $I_{\rm b}$ flowing through the junction as
\be
E_{\rm J}=\sqrt{{E_{\rm J0}}^2-(I_{\rm b}/(2e))^2}-{\gamma'}^2I_{\rm b}/(2e),
\ee
where $E_{\rm J0}$ is the Josephson energy in the absence of a bias current,  and
$\gamma'$ is the Josephson phase  measured from the potential bottom position. 
The potential barrier against the switching of the junction is approximately given by
$
\Delta V\simeq \frac{3}{\sqrt{2}}E_{\rm J0}x^{3/2}(1-x/24),
$
where $x=I_{\rm b}/(2e E_{\rm J0})$.
Therefore, we can approximately regard the circuit as a tunable LC resonator
 (a harmonic oscillator) under the conditions
$
{\gamma'}^2 \ll 2e E_{\rm J}/I_{\rm b}
$
and
$
k_{\rm B} T, E_{\rm noise} \ll \Delta V.
$ 
The former condition is necessary in order to neglect the nonlinearity of the potential and 
it restricts the amplitude of the resonator.
The latter prevents unwanted excitations, {\it e.g}., those due to
thermal,  noise, from causing switching.
As a result, by changing the bias current, we can tune the resonant frequency of a Josephson junction
 resonator circuit from 70 \% to 100 \% of that in an unbiased resonator.

Suppose that there are a number of flux qubits ($i=1,2,\ldots$) in a region of sub-millimeter size.
The level splitting of each qubit is $\omega_i=\sqrt{{\varepsilon_i}^2+{\Delta_i}^2}$, 
where $\varepsilon_i$ can be controlled by changing 
the external magnetic flux piercing the qubit ring, and $\Delta_i$ is the tunneling matrix element of the qubit.
In superconducting flux qubit systems, level splitting $\omega_i$ between two states is 
usually different  for each qubit. 
All qubits interact with the same Josephson resonator discussed above via mutual inductance $M_i$ ($i=1,2,\ldots$),
as  illustrated in Fig. \ref{Layout}.
When we represent the qubits with Pauli matrices and the resonator with creation and annihilation operators, 
we obtain the Hamiltonian
\bea
H=\sum_i \frac{\omega_i}{2}\sigma_{iz} +\omega_a \left(a^{\dagger}a+\frac1{2}\right) \nn
+\sum_i g_i (a^{\dagger}+a)(\cos\frac{\theta_i}{2}\sigma_{iz}-\sin\frac{\theta_i}{2}\sigma_{ix}) \nn 
-\sum_i (\cos\frac{\theta_i}{2}\sigma_{iz}-\sin\frac{\theta_i}{2}\sigma_{ix})f_i \cos\omega_{\rm ex} t,
\eea
where $\tan\theta_i=\Delta_i/\varepsilon_i$, and $g_i \propto M_i$ is the qubit-resonator coupling constant.
Here, the last term corresponds to microwave irradiation to qubits with the frequency $\omega_{\rm ex}$.

Now we choose two qubits $i=1,2$ and consider the coupling between them.
First, we tune the frequency of the resonator to $\omega_{\rm r}=\frac1{2}(\omega_1+\omega_2)+\delta$. 
Then, 
we apply a microwave of the frequency $\omega_{\rm ex}=\frac1{2}(\omega_2-\omega_1)+\delta_{\rm ex}$. 

The energy diagram of the two-photon Rabi oscillation caused by the microwave irradiation is illustrated in 
Fig. \ref{F:diagram}  $^8$. 
Here, the notation of the state, $|g,\ e,\ n\rangle$ means that qubit 1 is in its ground state and  qubit 2 is in 
its excited state, and the resonator is in the boson number state of $|n\rangle$. 
When we expect the transition between $|e,\ g,\ n\rangle$ and $|g,\ e,\ n\rangle$, the intermediate state is
$|g,\ g,\ n+1\rangle$ or $|e,\ e,\ n-1\rangle$.  When the detuning $\delta$ is  
exactly equal to 0, the energies of these intermediate states
are placed exactly at the middle between $|g,\ e,\ n\rangle$ and $|g,\ e,\ n\rangle$. 
Then, the irradiation resonates to the single photon transition to/from the
intermediate states and real transitions occur. By applying a finite detuning $\delta$, such real transitions are suppressed and
the two-photon  transitions between $|e,\ g,\ n\rangle$ and $|g,\ e,\ n\rangle$ are induced.   
In the absence
of the resonator, the transition is negligibly small.

Next, we show the results of numerical calculations. 
For simplicity, parameter values were fixed as follows: 
$\omega_1=1.0$, $\omega_2=1.2$, $\theta_1=\theta_2=\pi/6$, $g_1=0.05$, $g_2=0.06$, $f_1=0.02$ and $f_2=0.03$.
Moreover, the irradiation frequency was set as $\omega_{\rm ex}=\frac1{2}(\omega_2-\omega_1)+\delta_{\rm ex}$.  
The detuning $\delta_{\rm ex}=0.0074$ was applied in order to  resonate the true energy splitting affected by the coupling with the resonator.
The resonator frequency $\omega_{\rm r}=\frac1{2}(\omega_1+\omega_2)+\delta$ with $\delta$ from -0.1 to +0.1 was examined. 
The time evolution of the density operator of the two-qubit-resonator composite system were calculated by solving
the master equation 
$\frac{\id}{\id t}\rho=\frac1{i}[H,\rho]+\frac{\Gamma}{2}(2 a\rho a^{\dagger}-a^{\dagger}a \rho-\rho a^{\dagger}a)$
with the backward-Euler method. This master equation corresponds to the case where linear loss in the resonator is considered.

Figure \ref{F:TEP} shows the time evolution of the system without any decoherence, i.e., $\Gamma=0$.
The detuning of the resonator is $\delta=-0.02$. The initial state is  $|e, g, 0\rangle$. 
In Fig. \ref{F:TEP}(a),  we can see the population of the initial state is  transfered to the $|g, e, 0\rangle$.
This is the Rabi oscillation of $|e,\ g,\ 0\rangle \leftrightarrow |g,\ e,\ 0\rangle$, which forms an entangled state
$|\psi\rangle=\alpha |e,\ g,\ 0\rangle +e^{i \varphi}\sqrt{1-|\alpha|^2} |g,\ e,\ 0\rangle$.

To confirm the entanglement, we show the time evolution of the concurrence $C$ of the qubits in Fig. \ref{F:TEP}(b).
The concurrence is a measure of the extent of entanglement and defined by
$C=\sqrt{\lambda_1}-\sum_{k>1}\sqrt{\lambda_k}$, where $\lambda_1\geq\lambda_2\geq\lambda_3\geq \ldots$,
and $\lambda_k$ is an eigenvalue of  $\rho_y\rho_y^*$ with the density operator $\rho_y=\sigma_{1y}\sigma_{2y}\rho$.

$C=1$ means that a maximally entangled qubit pair is formed.  In our calculation, $C$ approaches   unity,
although a rapid oscillation is superposed.
The rapid oscillation comes from the real excitation of the state $|g,\ g,\ 1\rangle$.
This type of real excitation is much suppressed when the microwave irradiation is weak.
However, under weak irraditaion, the period of the two-photon Rabi oscillation becomes long
resulting in taking a long time for entangling two qubits. As discussed below,
entanglement formation should be completed within the coherence time. Therefore,
weak irraditation is not appropriate. Some optimization, such as detuning of the resomator frequency
is still possible to suppress unwanted real excitation. 
 
We also examined the influence of decoherence on the behavior of the coupler. 

An almost maximal entanglement can be formed when we use a resonator with the Q value of more than one thousand.
The formation takes a few hundred times  the period of the Larmor rotations of qubits, $2 \pi/\omega_i$.
For a qubit of $\omega_i\sim 4 \ \mathrm{GHz}$, this approximately  corresponds to 100 nanoseconds. 
Therefore, a controlled-NOT gate containing approximately ten  two-qubit operations, can be constructed 
when the coherence time of qubits exceeds  microseconds. This condition can be met in the near future.

It is worth  mentioning that this coupler can  make an interaction 
between remote qubits because they interact with the same
resonator whose size can become the order of sub-millimeter. 
This makes   the entanglement formation between remote qubits considerably fast
compared to that using successive nearest neighbor interactions. 
 
The resonant conditions  
are rather strict. When the  frequencies deviate by a few percent from the conditions, two-photon transition becomes 
negligibly small. 
This strictness offers a good selectivity of the target pair of qubits from other qubits.
We can switch the target qubit-pair within 
a few nanoseconds by changing the DC-bias and the irradiation frequency.

In summary, we proposed a qubit-qubit coupler for superconducting flux qubits,
using  a Josephson resonator and a microwave irradiation. 
The coupler can make an interaction
between  two remote qubits among many qubits. We have 
already performed a quantum interaction behavior of a composite system 
consisting of a flux-qubit and a superconducting
LC resonator $^9$. 
Moreover, we have succeeded in controlling  two-photon Rabi oscillations 
in single flux-qubit systems $^{10}$. 
Therefore,
we have the elementary experimental techniques necessary to realize the coupler discussed here. 

\vspace{1em}

This work was partially supported by CREST-JST.

\vspace{1em}

\newpage 

{\Large\bf References}

\setlength{\parindent}{0mm}

$^1$Y. Makhlin, A. Shnirman and G. Sch\"{o}n, Rev. Mod. Phys., {\bf 73},  357 (2001).

$^2$ A. Blais, A. Massan van den Brink and A. M. Zagoskin,
Phys. Rev. Lett., {\bf 90}, 127901 (2003).

$^3$
J. Q. You, J. S. Tsai, and Franco Nori, 
 Phys. Rev. {\bf  B 68}, 024510 (2003).

$^4$ B. L. T. Plourde, J. Zhang, K. B. Whaley, F. K. Wilhelm, T. L. Robertson, T. Hime, S. Linzen, P. A. Reichardt, C.-E. Wu, and John Clarke,
Physi. Rev. {\bf B 70}, 140501(R)  (2004).

$^5$ P. Bertet, C. J. P. M. Harmans, and J. E. Mooij, Phys. Rev. {\bf B 73} , 064512 (2006).

$^6$ A. O. Niskanen, K. Harrabi, F. Yoshihara, Y. Nakamura, and J. S. Tsai, Phys. Rev. {\bf B 73}, 094506 (2006).

$^7$ J. I. Cirac and P. Zoller, Phys. Rev. Lett., {\bf 74}, 4091 (1995).

$^8$ A qubit-qubit coupler for a trapped ion system using similar two-photon Rabi oscillation was proposed in,
A. S\o rensen and K. M\o lmer,  Phys. Rev. Lett. {\bf 82}, 1971 (1998). However, their resonator is not tunable and the energy diagram
is different.

$^9$ J. Johansson, S. Saito, T. Meno, H. Nakano, M. Ueda, K. Semba, and H. Takayanagi,  Phys. Rev. Lett. {\bf 96}, 127006  (2006).

$^{10}$ S. Saito, T. Meno, M. Ueda, H. Tanaka, K. Semba, and H. Takayanagi,  Phys. Rev. Lett.  {\bf 96},  107001 (2006).

\newpage
{\Large\bf Figure Captions}

\setlength{\baselineskip}{8mm}

\begin{figure}[ht]
\setlength{\baselineskip}{8mm}
\includegraphics[width=9cm]{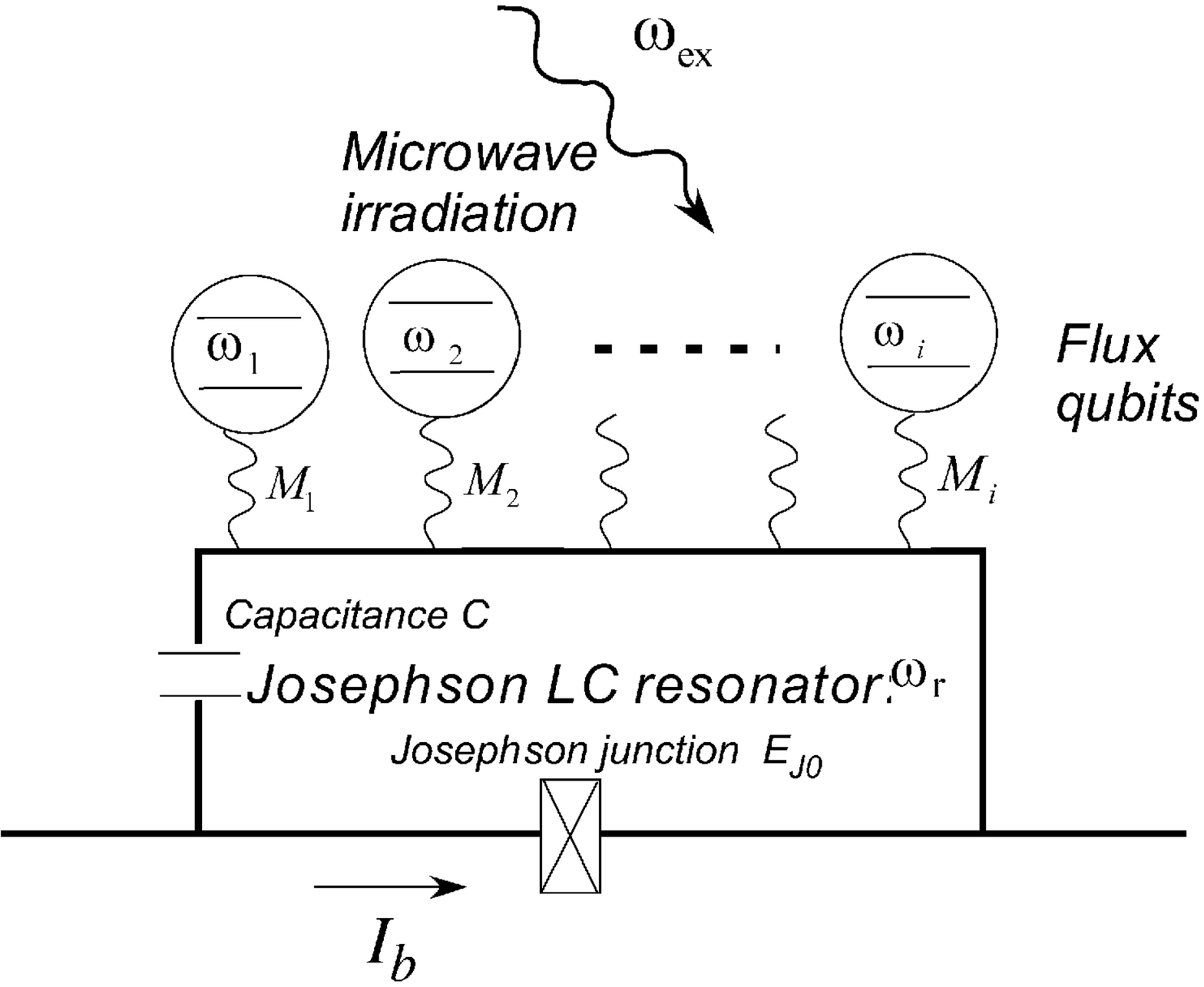}
\caption{\label{Layout}}
Flux qubits interacting with a common Josephson resonator. The resonance frequency
of the resonator can be tuned by changing the bias current $I_{\rm b}$. Each flux qubit with 
resonant frequency $\omega_i $ couples to the resonator
via mutual inductance $M_i$. We apply a microwave of frequency 
$\omega_{\rm ex}\simeq (\omega_i\pm \omega_j)$ to make entanglement between the qubits $i$ and $j$.
\end{figure}

\begin{figure}
\setlength{\baselineskip}{8mm}
\includegraphics[width=8cm]{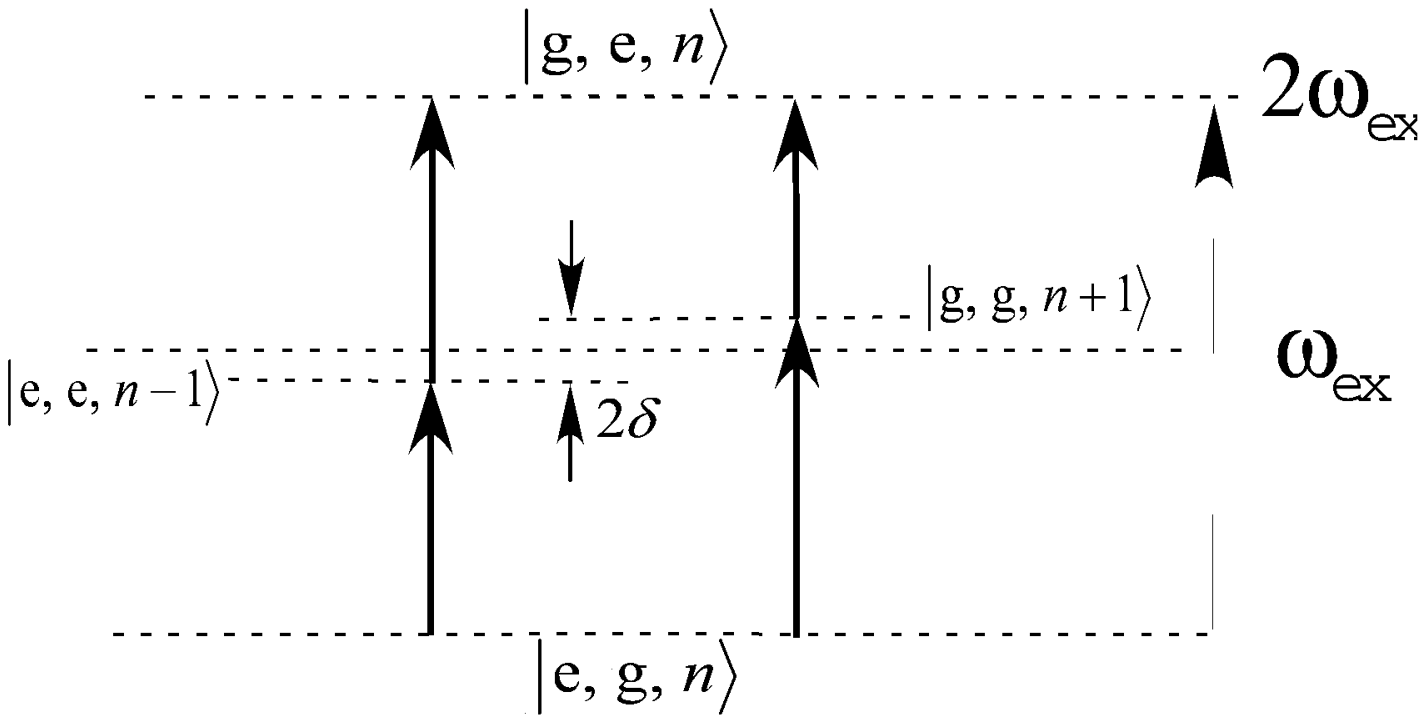}
\caption{
\label{F:diagram}
}
Schematic energy diagram for the two-photon Rabi oscillation in the two-qubit-resonator system. 
The Rabi oscillation causes coherent transition between $|e, g, n\rangle$ and $|g, e, n\rangle$.
The energy splitting between these two states corresponds to two times  the microwave frequency $\omega_{\rm ex}$.
This is a nonlinear (second-order) transition. Energies of the  intermediate  states 
$|e, e, n-1\rangle$ and $|g, g, n+1 \rangle$ are shifted to approximately the middle between
$|e, g, n\rangle$ and $|g, e, n\rangle$ by the interaction with the resonator, resulting
in enhancement of the nonlinear (second-order) transition.
\end{figure}

\begin{figure}
\setlength{\baselineskip}{8mm}
\includegraphics[width=10cm]{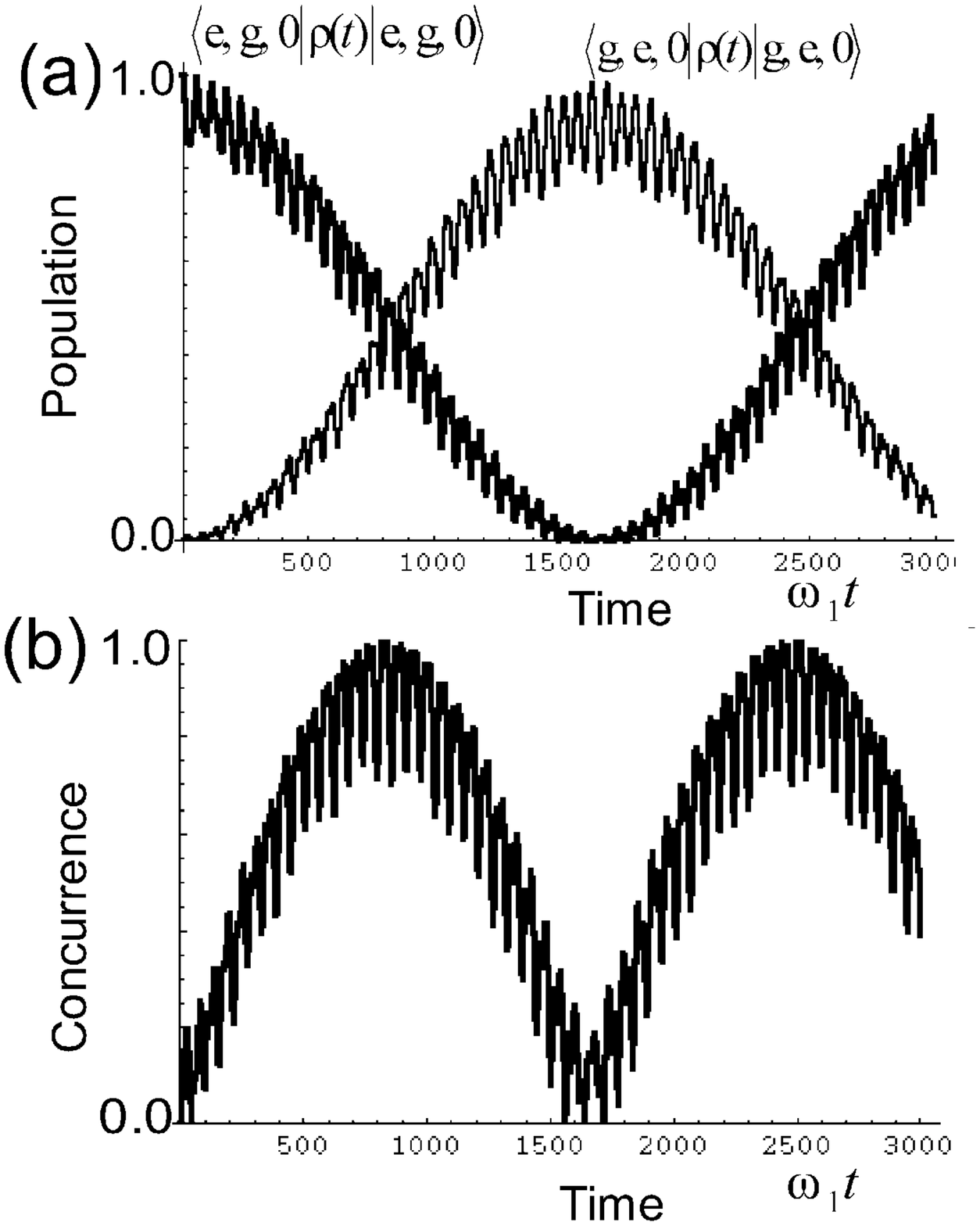}
\caption{\label{F:TEP}
}
Time evolution caused by microwave irradiation, where, there is no decoherence in the system.
The initial state is $|e, g, 0\rangle$. Rabi oscillation makes a superpostion 
of $|e, g, 0\rangle$ and $|g, e, 0\rangle$, which is an entangled state. (a) populations of the two states. 
(b) concurrence of the entangled state. 
\end{figure}

\begin{figure}
\setlength{\baselineskip}{8mm}
\includegraphics[width=10cm]{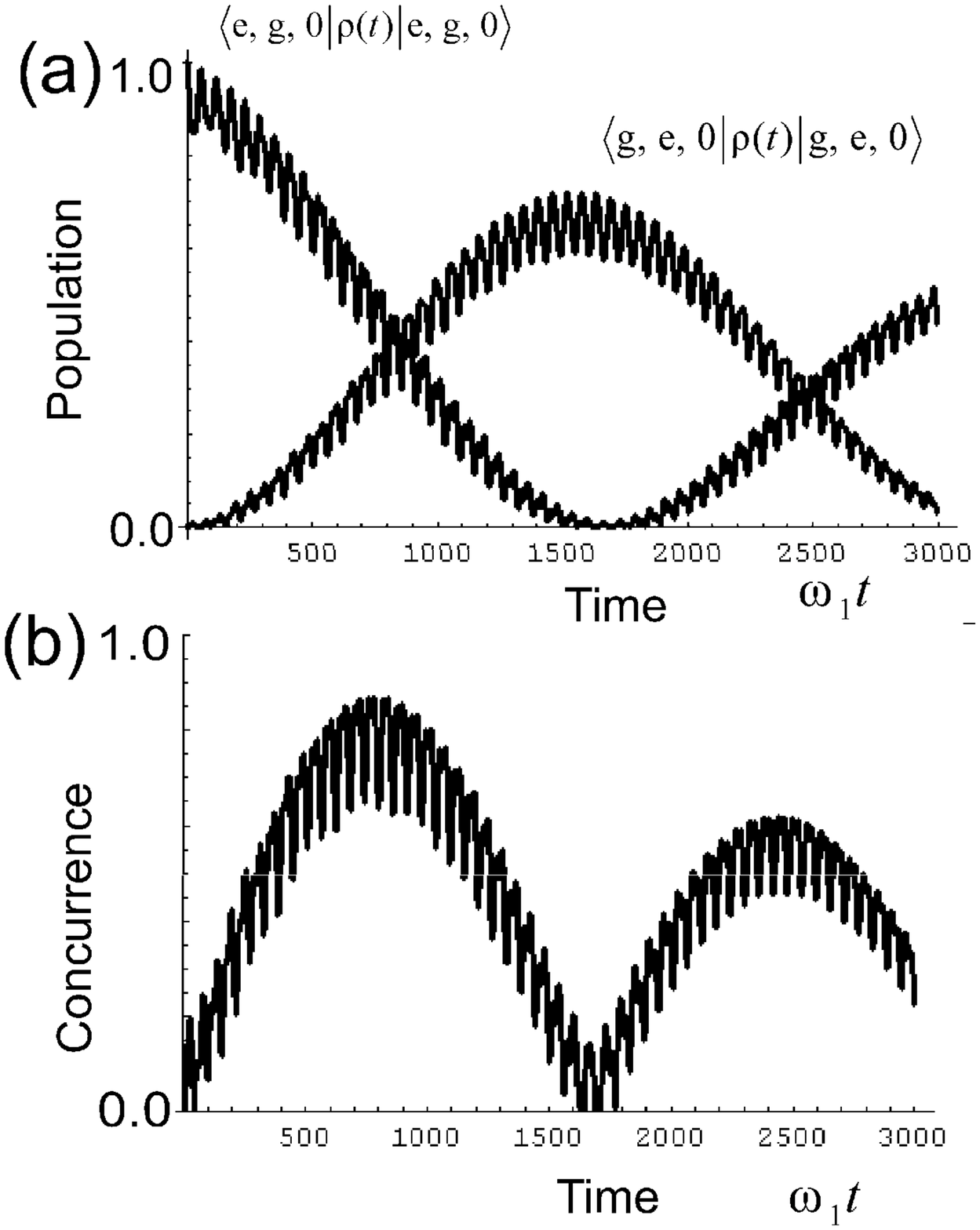}
\caption{\label{F:TED}}
Time evolution caused by maicrowave irradiation, with decohererence ($\Gamma=0.002$). Linear loss in the resonator
is provided. There is no direct decoherence to qubits. (a) populations in the two states. (b) concurrence
of entanglement. Entanglement is much poluted by decoherence.
\end{figure}

\end{document}